\documentstyle[prl,aps,preprint]{revtex}
\tightenlines
\input{epsf}
\def\mathbf{\bf}

\begin{document}

\title{A Fermi Surface study of Ba$_{1-x}$K$_{x}$BiO$_{3}$} 

\author{
S. Sahrakorpi$^{1}$, B. Barbiellini$^{2}$,
R. S. Markiewicz$^{2}$, 
S. Kaprzyk$^{2,3}$,
M. Lindroos$^{1,2}$,
A. Bansil$^{2}$
}

\address{$^{1}$ Institute of Physics, Tampere University of Technology,
33101 Tampere, Finland}

\address{$^{2}$Department of Physics, Northeastern University, Boston, 
MA 02115 USA}

\address{$^{3}$Academy of 
Mining and Metallurgy AGH, 30059 Krak\'{o}w, Poland} 

\date{\today}
\maketitle
\begin{abstract}
We present all electron computations of the 3D Fermi surfaces (FS's) in 
Ba$_{1-x}$K$_{x}$BiO$_{3}$ for a number of different compositions 
based on the selfconsistent 
Korringa-Kohn-Rostoker coherent-potential-approximation (KKR-CPA) approach 
for incorporating the effects of Ba/K substitution. By assuming a simple
cubic structure throughout the composition range, the evolution of the
nesting and other features of the FS of the underlying 
pristine phase is correlated with the onset of various structural 
transitions with K doping. A parameterized scheme for obtaining an 
accurate 3D map of the FS in Ba$_{1-x}$K$_{x}$BiO$_{3}$ for an arbitrary doping 
level is developed. We remark on the puzzling differences between the phase 
diagrams of Ba$_{1-x}$K$_{x}$BiO$_{3}$ and BaPb$_{x}$Bi$_{1-x}$O$_{3}$
by comparing aspects of their 
electronic structures and those of the end compounds BaBiO$_{3}$, 
KBiO$_3$ and BaPbO$_3$. Our theoretically predicted FS's in the cubic phase
are relevant for analyzing high-resolution Compton scattering and 
positron-annihilation experiments sensitive to the electron momentum 
density, and are thus amenable to substantial experimental 
verification. 

\end{abstract}


\section{Introduction}

The cubic perovskite Ba$_{1-x}$K$_{x}$BiO$_{3}$\cite{gyorgy,cava} 
which achieves a maximum 
transition temperature of 32K (for $x\approx 0.4$) has been the subject of 
numerous studies. Despite some similarity to the better known high-T$_c$ 
cuprates, the system is three dimensional and lacks strong magnetic 
properties in the normal state. The vibrational breathing mode of the 
BiO$_6$ octahedra appears to yield a strong electron-phonon coupling 
\cite{huang,liech,meregalli,hamada} which together with
dielectric effects may explain superconducting properties \cite{pwp,dayan}.
However, recently observed anomalous temperature dependencies of the 
critical magnetic fields and vanishing discontinuities in the specific heat 
and magnetic susceptibility suggest a fourth order transition to 
superconductivity \cite{kumar}. Therefore, in contrast to the standard 
BCS picture, thermodynamic properties seem almost unchanged through the 
transition. 

Ba$_{1-x}$K$_{x}$BiO$_{3}$ also possesses 
a rich structural phase diagram\cite{baumert,Pei} 
as a function of K doping. 
In the range $0<x<0.12$, the system assumes a 
monoclinic structure which can be obtained from the cubic structure via 
small tilting and breathing distortions of the BiO$_6$ octahedra; 
in the undoped compound the tilting angle along 
$(1,1,0)$ is estimated to be $11.2^{\circ}$ and 
the breathing distortion to be $0.085$ $\rm\AA$. For 
$0.12 < x < 0.37$, the structure is orthorhombic, admitting tilting but 
not breathing distortion. Finally, for $0.37 < x <0.53$, when the cubic 
phase is stabilized, the system becomes metallic.
Although it is widely believed that the insulating phases 
for $x < 0.37$ are caused by charge density instabilities associated with
the breathing and tilting distortions, it has proven
difficult to establish this in terms of first principles computations. 
Very recent total energy calculations on distorted lattices \cite{meregalli} 
indicate that, in contrast to earlier results\cite{liech},
the LDA substantially underestimates the size of the breathing distortion 
and yields a metallic ground state. Perhaps 
correlation corrections beyond the LDA are necessary in
order to explain the insulating phases. 

In this article, we report highly accurate, all electron computations 
of 3D Fermi surfaces in Ba$_{1-x}$K$_{x}$BiO$_{3}$
for a number of different compositions. 
All calculations pertain to the simple cubic (SC) lattice and 
are parameter free except for the use of the Korringa-Kohn-Rostoker coherent 
potential approximation (KKR-CPA) to treat the effects of Ba/K substitution, 
and the local density approximation (LDA) for treating 
exchange-correlation effects. Our motivation for invoking the SC 
structure throughout the composition range is that in this way we are 
in a position to focus on the evolution of nesting and other features 
of the Fermi surface (FS) in the underlying pristine phase and to 
delineate how the appearance of such features correlates with the 
onset of various structural transitions with K doping. Note that 
lattice distortions in Ba$_{1-x}$K$_{x}$BiO$_{3}$ are relatively small, 
and pseudo-cubic lattice parameters are easily assigned in all cases. 

Our computations show clearly that the highest occupied band in 
BaBiO$_{3}$, in which the FS resides, remains virtually unchanged in shape 
upon substituting Ba with K, and that the associated states near the 
Fermi energy ($E_F$) continue to possess long lifetimes since they 
suffer little disorder induced scattering in the alloy. 
This circumstance allows us to fit this band in BaBiO$_{3}$
in terms of a Fourier-like expansion which accurately describes 
the highest occupied band in Ba$_{1-x}$K$_{x}$BiO$_{3}$
for {\it all compositions x}; 
a knowledge of the $E_F$ then yields the corresponding FS. In this 
way, we provide a useful parametrized form which permits a 
straightforward determination of the full 3D Fermi surface in 
cubic Ba$_{1-x}$K$_{x}$BiO$_{3}$ for any arbitrary K doping level. 

Highlights of some of the issues addressed, together with an outline of 
this article are as follows. Section II gives an overview of the methodology
and provides associated technical details of the computations. The 
presentation of results in Section III is subdivided into several 
subsections. Subsection IIIA discusses changes in topology of the FS
with K doping and attempts to correlate these changes with the 
observed structural transformations in Ba$_{1-x}$K$_{x}$BiO$_{3}$
invoking Hume-Rothery 
and Van Hove-Jahn Teller scenarios. Subsection IIIB takes up the 
question of parametrizing the FS, and gives details of the parameters 
which describe the doping dependent FS of cubic Ba$_{1-x}$K$_{x}$BiO$_{3}$. 
Subsection IIIC compares aspects of 
the electronic structures of Ba$_{1-x}$K$_{x}$BiO$_{3}$
and BaPb$_{x}$Bi$_{1-x}$O$_{3}$
as well as those of the end compounds BaBiO$_3$, KBiO$_3$
and BaPbO$_3$ with an eye towards understanding some puzzling 
differences between the phase diagrams of Ba$_{1-x}$K$_{x}$BiO$_{3}$
and BaPb$_{x}$Bi$_{1-x}$O$_{3}$. 
Subsection IIID discusses how our theoretical FS's for the cubic 
phase are relevant for analyzing experiments sensitive to the 
momentum density of the electron gas (positron annihilation, 
high resolution Compton scattering), and are thus amenable to substantial 
experimental verification; a recent ARPES measurement of doping 
dependence of the chemical potential in Ba$_{1-x}$K$_{x}$BiO$_{3}$
is also discussed 
in order to gain insight into the band renormalization at the 
Fermi energy. Section IV summarizes our conclusions. 
Finally, concerning related work, it may be noted that we are not 
aware of a systematic study of the evolution of the FS of 
Ba$_{1-x}$K$_{x}$BiO$_{3}$
with K doping in the literature, although aspects of the problem have 
been commented upon by various authors 
\cite{meregalli,hamada,mattheis83,mattheis88}.

\section{Overview of methodology, computational details}

Before proceeding with the computation of the FS for a given K doping $x$, 
we first obtained the charge selfconsistent KKR-CPA crystal potential in 
Ba$_{1-x}$K$_{x}$BiO$_{3}$ assuming random substitution of Ba by K; for 
details of our KKR-CPA methodology, we refer to Refs. 
\cite{kkrcpa,kap90,ban92}. The charge as well as the KKR-CPA selfconsistency 
cycles have been carried out to a high degree of convergence in all cases; 
for example, the final Fermi energies are accurate to about $2$ mRy and the 
total charge within each of the muffin-tin spheres to about $10^{-3}$ 
electrons. The total energies were not minimized to determine the lattice 
constants. The experimental lattice data was used instead, but otherwise the 
computations are parameter free. The simple cubic lattice constants used for 
BaBiO$_3$ and KBiO$_3$ are: $4.3485$ $\rm\AA$ and $4.2886$ $\rm\AA$ 
respectively \cite{Pei}. For intermediate compositions, lattice constants 
were obtained via Vegard's Law. The muffin-tin radii of Bi and O were taken 
to be $a/4$, where $a$ denotes the composition dependent lattice constant. 
The radius of the Ba or K sphere (recall that within the KKR-CPA scheme the 
two radii must be equal as these atoms occupy the same site 
randomly) was chosen by requiring the Ba/K sphere to touch 
the O-sphere, which gives the value $(1/\sqrt{2}-1/4) a$ 
for the Ba/K radius.\cite{kkrcpa} The aforementioned choices of the radii 
provide a good convergence of the crystal potential, and in any event, the 
results are not sensitive to these details. The calculations employ 
the Barth-Hedin exchange-correlation functional\cite{barth72} and are 
semi-relativistic with respect to the valence states, but the core states 
are treated relativistically. 
However, the relativistic effects on the valence states are expected 
to be small. In particular, the band giving rise to the FS is 
built mainly from the Bi-6s and O-2p orbitals which are affected 
little by the spin-orbit coupling; the effect of Bi $6$ $p$ 
admixture on the bands is estimated to be on the order of 
$0.1$ eV.\cite{mattheis83}
The maximum $\ell$-cut-offs used are 
$\ell_{max}=3$ for Ba, K and Bi-sites, and $\ell_{max}=2$ for O-atoms.

Once the selfconsistent crystal potential is determined using the 
preceding procedure, the Fermi surface in a disordered alloy 
is computed by evaluating the spectral density function
$A({\mathbf{p}},E)=-(1/{\pi})\mbox{Im}(G({\mathbf{p}},E))$,
where $G({\mathbf{p}},E)$ is the one-particle ensemble averaged 
KKR-CPA Green function at a given momentum $\mathbf{p}$ 
and energy E. The radius 
of the FS along a given direction $\bf {\hat p}$ is 
then defined by the position of the peak in 
$A(p,E)$ at $E=E_F$; the finite width of the spectral peaks reflects
the disorder induced scattering of states, and would in general yield 
a FS in an alloy which is smeared or blurred.\cite{bansil93,bansil82}
In the present case of Ba$_{1-x}$K$_{x}$BiO$_{3}$, 
however, it turns out that 
the Bi-O states near the $E_F$ are virtually unaffected by Ba/K 
substitution and therefore suffer little damping ($\lesssim 1$ mRy). 
For this reason, the KKR-CPA variations in the $E_F$ 
in Ba$_{1-x}$K$_{x}$BiO$_{3}$ are also 
close to the rigid band values based on the BaBiO$_{3}$ band structure. 
In order to obtain highly accurate 3D maps of the FS discussed below, 
a uniform net of about $10^5$ $\mathbf{k}$-points in the 
irreducible Brillouin zone has been employed. 
Finally, we note that the specific parameters used in the 
density of states and related computations on 
BaPb$_{x}$Bi$_{1-x}$O$_{3}$ presented in 
this work are similar to those detailed above; the lattice 
constant of BaPbO$_3$ was $4.2656$$\rm\AA$ \cite{cox}.

\section{Results and discussion}

\subsection{Evolution of the Fermi Surface with Doping, Structural 
Transitions}

Figures 1-4 present 3D images of the FS in the cubic phase for 
K concentrations $x$ = 0.67, 0.40, 0.13 and 0.0, together with 
three different cross-sections in the (001) and (110) planes. 
With reference to these figures 
we will discuss how nesting features evolve and correlate 
with the occurrence of structural transitions 
in Ba$_{1-x}$K$_{x}$BiO$_{3}$ with K doping. 

The FS for $x=0.67$ is shown in Fig. \ref{FS_cub}. The composition is 
at the upper limit of stability ($x\approx 0.6 - 0.7$) of the cubic phase. 
The FS is a flattened free-electron-like sphere which appears nearly cubic in 
shape. This is evident in the 3D rendition of Fig. 1(a) as well as 
in the squarish appearance of sections of Figs. 1(b)-1(c). 
Our computations indicate that the FS becomes even more cube-like 
for $x> 0.67$ (not shown); since the cubic shape is particularly 
susceptible to nesting, one may speculate a connection with the 
aforementioned phase stability limit. Incidentally, asphericity of the 
FS introduces momentum dependence in the Eliashberg equation
with subtle consequences for superconducting properties\cite{weger}.

With decreasing $x$, the FS grows in size as seen in Fig. \ref{FS_htc} 
for $x$=0.40; this composition has been chosen to lie close to the 
cubic-orthorhombic phase boundary at $x$=0.37. Since the 
orthorhombic unit cell is very similar to FCC \cite{mattheis83}, the
associated Brillouin zone (BZ) is also drawn in Fig. 2. The FS is seen 
to make contact with the hexagonal face of the bcc zone; this is more 
clear in the (110)-section of Fig. 2(d). These results suggest that 
the cubic-orthorhombic transition may be viewed as a Hume-Rothery type
structural instability\cite{HR} to some larger unit cell which arises when the 
FS crosses the BZ of the associated supercell. We find that 
the FS becomes tangent to the hexagonal face at $x=0.45$. 
Note that the Hume-Rothery rules require the transition to occur not at the 
point of first contact with the BZ, but after the FS has grown to slightly 
overlap the zone boundary\cite{Bland}.  
In the present case these arguments would thus predict 
a transition to an FCC structure at $x\approx 0.4$ where the cubic FS 
has already broken through the zone boundary. Recall that the
orthorhombic structure involves lattice distortions via tilting mode 
phonons with wavevector ${\bf R}=(1,1,1)\pi /a$ 
\cite{liech,meregalli}, which is 
consistent with Fig. 2(d) where the spanning vector (denoted by the arrow) 
is indeed seen to be approximately equal to ${\bf R}$.

It is noteworthy that different Hume-Rothery phases presumably involve 
a succession of free energy minima as a function of composition\cite{Bland}. 
If so, there is the possibility that the system will actually go into 
a mixed phase at the transition, similar to the mixed $\alpha$ plus $\beta$ 
phase in brasses\cite{hansen}. 
In the Ba$_{1-x}$K$_{x}$BiO$_{3}$ system, an 
incommensurate modulation has been
observed by electron but not neutron diffraction\cite{Pei},
suggestive of a fluctuating or nanoscale phase separation, which is 
reminiscent of the stripe-like phases found in the cuprates and 
related oxides.

We consider Fig. \ref{FS_vhs} next for $x=0.13$ 
where Ba$_{1-x}$K$_{x}$BiO$_{3}$ undergoes the orthorhombic
to monoclinic transition. Compared to $x=0.40$ (Fig. 2), the FS has become 
more rounded. The most striking feature however is that the FS has grown to 
just begin making contact with the zone boundary at the X-point. 
In fact, the band structure of cubic BaBiO$_{3}$(see subsection C below)
contains a saddle point at X which lies approximately 0.1 eV below 
the Fermi energy. 
The associated van Hove singularity (VHS) in the density of states crosses 
the Fermi level around $x\approx 0.13$ and gives rise to the change in the FS 
topology seen in Fig. 3. Computations of Ref. \cite{liech} indicate 
that the electron-phonon coupling parameter $\lambda$ can increase 
sharply as $x$ decreases below 0.13 causing the breathing mode phonon to
become unstable. 

It is interesting to ask the question: Since the orthorhombic and 
monoclinic distortions possess nearly the same FCC BZ, what is the 
driving force behind the transformation at $x=0.13$?  The FS of Fig. 3 
suggests an interpretation in terms of a Van Hove-Jahn-Teller 
scenario\cite{rm8b,RiSc}. When the VHS intersects the Fermi energy, 
there are three independent VHS's whose degeneracy cannot be lifted 
in the orthorhombic phase. Since each VHS involves a substantial 
density of states, the system can gain energy via a Jahn-Teller 
distortion to a lower symmetry phase (such as monoclinic) which 
lifts the degeneracy between the VHS's. 

Fig. \ref{FS_pure} shows the FS of cubic BaBiO$_{3}$ for completeness. 
The FS is a distorted sphere with large necks at X. No significant 
changes in the topology of the FS take place over the 
composition range $0<x<0.13$.

\subsection{A Parameterized form for the doping dependent Fermi surface }

We fit first the Bi6s-O2p band $E(k_x,k_y,k_z)$ in BaBiO$_{3}$
(see Fig. \ref{ban}), which 
gives rise to the FS, in terms of the following Fourier-like expansion:
\begin{eqnarray}
E(k_x,k_y,k_z) &= &E_0 +t_1( X + Y + Z) 
+ t_2(X Y + X Z + Y Z)+ t_3 X Y Z \\
\nonumber
& & + t_4( X_2 + Y_2 + Z_2) 
+ t_5(X Y_2 + X Z_2 + Y Z_2 + X_2Y + X_2 Z + Y_2 Z) \\
\nonumber
& &  + t_6(X_2 Y_2 + X_2 Z_2 + Y_2 Z_2)~,
\end{eqnarray} 
where $X=\cos(k_xa)$, $Y=\cos(k_ya)$, $Z=\cos(k_za)$,
$X_2=\cos(2k_xa)$, $Y_2=\cos(2k_ya)$, $Z_2=\cos(2k_za)$; 
$a$ is the lattice constant and $E_0$ is the average band energy. 
The k-dependence on the right side of Eq. (1) possesses the form of a 
tight-binding band and in this sense $t_n$ may be viewed as the $n$th nearest
neighbor "hopping integral". The values of various 
parameters which fit the computed 3D band 
are (in eV): $E_0=-0.288$ (with respect the Fermi level of BaBiO$_{3}$),
$t_1=-0.6191$, $t_2=-0.4313$, $t_3=0.0816$, $t_4=0.1034$, 
$t_5=0.1361$ and $t_6=-0.0449$. Higher order terms in the expansion 
are found to be negligibly small. The fit is valid throughout 
the composition range in Ba$_{1-x}$K$_{x}$BiO$_{3}$ since, 
as already noted, the 
Bi6s-O2p band remains essentially unchanged in shape near the $E_F$ 
with K/Ba substitution. The fact that the terms with $t_3-t_6$ are 
significant in obtaining an accurate fit indicates that the 
associated interaction parameters possess a fairly long range.
Incidentally, supercell simulations indicate 
that electronic states near the Fermi level are not
sensitive to short-range ordering effects
\cite{mattheis88}.

The constant-energy surface can now 
be obtained for any given value of the energy by solving Eq. (1); 
in order to obtain the FS at a given doping, we need only specify the 
corresponding value of the $E_F$. For this purpose, we have parametrized 
the KKR-CPA values of $E_F(x)$ in Ba$_{1-x}$K$_{x}$BiO$_{3}$ 
as a second-order polynomial: 
\begin{equation}
E_F(x)= a_1 x^2 + a_2 x~,
\end{equation}
where $E_F=0$ for $x=0$, $a_1 = -2.078$ eV, and $a_2 = -0.6612$ eV.
The solid lines in the sections of Figs. 1-4 show that Eqs. 1 and 2
provide an excellent fit to the 
3D Fermi surface in Ba$_{1-x}$K$_{x}$BiO$_{3}$
over the entire composition range. 
Notably, the positive sign of the ratio, $\tau=t_2/t_1$ reflects 
a concave down curving of the FS in the basal plane (see Figs. 1-4); 
in contrast, some cuprates possess FS's curving concave up\cite{mark}. 
For the special case where only the nearest neighbor hopping term 
$t_1$ is considered in Eq. (1), the 
FS will be perfectly nested at half filling, 
and hence unstable with respect to infinitesimal perturbations; 
the presence of interactions with farther out neighbors smears 
this singularity, although some softness in the system remains 
as already discussed above.

\subsection{Ba$_{1-x}$K$_{x}$BiO$_{3}$ vs. BaPb$_{x}$Bi$_{1-x}$O$_{3}$}
Despite substantial similarities, the phase diagrams of 
Ba$_{1-x}$K$_{x}$BiO$_{3}$ and 
BaPb$_{x}$Bi$_{1-x}$O$_{3}$ display significant differences. 
The monoclinic to 
orthorhombic transition occurs at roughly the same doping level in 
both systems, but the orthorhombic phase 
in BaPb$_{x}$Bi$_{1-x}$O$_{3}$ persists up 
to $0.6$ holes per band, and unlike Ba$_{1-x}$K$_{x}$BiO$_{3}$, 
it does not undergo 
the transition to the cubic phase. 
Since the FS's of Ba$_{1-x}$K$_{x}$BiO$_{3}$ and 
BaPb$_{x}$Bi$_{1-x}$O$_{3}$ may be expected to be roughly similar 
(in view of similarities 
of their electronic structures), on the face of it, the 
explanations of Section IIIA above 
for Ba$_{1-x}$K$_{x}$BiO$_{3}$ would appear to be 
applicable also to BaPb$_{x}$Bi$_{1-x}$O$_{3}$. 
Some insight into the puzzling behavior 
of BaPb$_{x}$Bi$_{1-x}$O$_{3}$ may be obtained 
by comparing the band structures near 
the Fermi energy of the end compounds BaBiO$_{3}$, 
KBiO$_3$ and BaPbO$_3$ shown in 
Fig. \ref{ban}, and the associated composition dependent densities of 
states in Ba$_{1-x}$K$_{x}$BiO$_{3}$ and BaPb$_{x}$Bi$_{1-x}$O$_{3}$
(Fig. \ref{dos}). 

The important point to note is that the band passing through the Fermi energy
in BaBiO$_{3}$ as well as KBiO$_3$ is a hybridized Bi-O band which is 
affected little when Ba is substituted by K in Ba$_{1-x}$K$_{x}$BiO$_{3}$; 
the associated density of states 
in Fig.~\ref{dos}a displays two distinct features near zero 
and $1.4$ eV (in BaBiO$_{3}$) which are weakly doping dependent. 
In sharp contrast, in
BaPb$_{x}$Bi$_{1-x}$O$_{3}$ the valence band changes from Bi-O to Pb-O, 
and the VHS's in the end compounds 
BaBiO$_{3}$ and BaPbO$_3$ lie around 
$-0.1$ eV and $2.0$ eV respectively, Fig. \ref{dos}b. [Structure from higher 
bands is evident above $2$ eV in Fig. \ref{dos}b.] 
As a result, the DOS of BaPb$_{x}$Bi$_{1-x}$O$_{3}$ alloys is 
characterized by a "split-band" 
behavior\cite{bansil93,bansil82}: when Bi is substituted by Pb, the 
VHS around $-0.1$ eV arising from the Bi-O band gradually loses 
spectral weight which gets transferred to the VHS around $2.0$ eV 
of the PbO band.  Consequently, states near the Fermi level will suffer 
substantial disorder induced scattering \cite{kkrcpa}.
The FS in BaPb$_{x}$Bi$_{1-x}$O$_{3}$ will then be quite smeared, 
rendering suspect the 
arguments of Section IIIA which assume a sharply defined FS.

\subsection{Comparison with experiments}
We note first that the theoretically predicted FS for the cubic phase
at $x=0.4$ (Fig. \ref{FS_cub}) is in good accord with the experimental 
FS deduced by Mosley {\em et al.} \cite{mosley} from positron annihilation 
measurements. As already mentioned in the introduction, the FS's computed 
for the cubic phase at compositions outside the range of stability of 
the phase (Figs. 1, 3 and 4) are nevertheless relevant for experiments, 
especially where one probes the momentum density of the electron gas. 
We elaborate on this point now. 
 
In a positron annihilation or high-resolution Compton 
scattering experiment 
\cite{bansil93,bansil82,mijnarends95}, the underlying spectral 
function involved is the 3D momentum density 
$\rho({\mathbf p})$ of the ground state. The FS signatures, 
which are scattered throughout the momentum space in $\rho({\mathbf p})$ 
can, in principle, \cite{footnote1} 
be enhanced by folding $\rho({\mathbf p})$ into the first BZ \cite{lcw} to 
obtain a direct map of the occupied states, i.e.
\begin{equation}
n({\mathbf k})=
\sum_{\mathbf G} \rho({\mathbf k+G}) ~,
\label{lcw}
\end{equation}
where $n({\mathbf k})$ is the occupation number for the Bloch state 
$\mathbf{k}$, and the summation extends over the set $\{\mathbf{G}\}$ 
of reciprocal lattice vectors. The FS may then be defined as the surface of 
maximum gradient of $n({\mathbf k})$ \cite{footnote2}. As already emphasized, 
the orthorhombic as well as 
the monoclinic phase of Ba$_{1-x}$K$_{x}$BiO$_{3}$
is derived via relatively small tilting and breathing 
distortions of BiO$_6$ octahedra. It will be sensible, therefore, 
to obtain $n({\mathbf k})$ from measured momentum densities 
by using vectors of the SC lattice in Eq. \ref{lcw} at all compositions 
of Ba$_{1-x}$K$_{x}$BiO$_{3}$. 
The evolution of the FS of Ba$_{1-x}$K$_{x}$BiO$_{3}$ with doping 
depicted in Figs. 1-4 should in this way be essentially verifiable 
experimentally despite the intervention of phase transitions. 
As the cubic symmetry is broken with doping and various 
gaps open up, the momentum density will be smeared over a range of 
approximately $E_{gap}/v_F$, where $E_{gap}$ is the energy gap 
and $v_F$ is the Fermi velocity  of the associated metallic 
state\cite{friedel}; this should, however, only produce 
relatively small modulations of $n({\mathbf k})$ based on 
the SC structure. In this vein, disorder effects in general 
yield a momentum smearing, $\Delta k =\gamma/v_F $ in terms 
of the disorder induced width (in energy) $\gamma$, although 
the value of $\gamma$ in Ba$_{1-x}$K$_{x}$BiO$_{3}$
is negligibly small at the Fermi energy.

An interesting recent experimental result concerns the shift in chemical 
potential $\mu(x)$ as a function of doping 
obtained by Kobayashi et al. \cite{xas} 
via XPS core level measurements in Ba$_{1-x}$K$_{x}$BiO$_{3}$. 
Fig. \ref{mu} shows that KKR-CPA predictions can be brought into 
line with the measurements provided the theoretical values 
are scaled down by a factor of $0.49$ (dashed curve), 
indicating that the dispersion of the quasiparticles near the 
Fermi energy may be given incorrectly in the underlying band 
structure. [The solid curve is the fit to the KKR-CPA values given by 
Eq. 2]. This is not surprising since it is well known that 
the excitation energies in general do not correspond
to the eigenvalues of the Kohn-Sham equation \cite{dft}.
Notably, Ref. \cite{xas} reports absence of any abrupt 
changes in the chemical potential through the orthorhombic 
and monoclinic phase transitions; however, any such jumps 
in $\mu(x)$ are expected to 
be small in light of the discussion of preceding sections, and 
lie presumably below the experimental resolution.
Also, core level shifts could be affected by crystal defects 
which may explain part of the discrepancy between theory and 
experiment \cite{lindau}.

\section{Summary and Conclusions}

We have obtained 3D Fermi surfaces in cubic 
Ba$_{1-x}$K$_{x}$BiO$_{3}$ over the entire 
composition range; representative results for $x=$ $0.67$, $0.4$, 
$0.13$ and $0.0$ are presented and discussed. 
The computations employ the selfconsistent 
KKR-CPA approach for treating the effects of Ba/K substitution within 
the framework of the local density approximation, 
but are parameter free otherwise. 
An examination of changes in the topology of 
the FS gives insight into transformations of the cubic phase into 
non-cubic structures as a function of K doping. Highlights of our 
specific conclusions are as follows: 

\begin{enumerate}
\item The cubic-orthorhombic
transition around $x=0.37$ is suggested to be a Hume-Rothery type 
instability when the FS makes contact with the BZ 
of the associated fcc lattice along the (111) directions. 
The orthorhombic-monoclinic transition around $x=0.13$ is interpreted 
within a van Hove- Jahn Teller scenario as the FS makes 
contact with the X-symmetry-point of the BZ. 

\item A parametrization scheme which allows an accurate determination 
of the 3D Fermi surface in 
cubic Ba$_{1-x}$K$_{x}$BiO$_{3}$ for an arbitrary doping level 
via a straightforward use of Eqs. 1 and 2 is developed. This scheme 
would be useful more generally for applications requiring FS integrals 
(e.g. response function computations) in Ba$_{1-x}$K$_{x}$BiO$_{3}$. 

\item We remark on the puzzling differences between the phase 
diagrams of Ba$_{1-x}$K$_{x}$BiO$_{3}$ and 
BaPb$_{x}$Bi$_{1-x}$O$_{3}$ by comparing the KKR-CPA electronic 
structures of Ba$_{1-x}$K$_{x}$BiO$_{3}$ and 
BaPb$_{x}$Bi$_{1-x}$O$_{3}$ and of the end compounds BaBiO$_{3}$,
KBiO$_3$ and BaPbO$_3$. The van Hove singularity in the highest occupied 
Bi-O band which is virtually unaffected by Ba/K substitution is found 
to be smeared strongly by Pb/Bi substitution, a fact which may be 
relevant in this connection. 

\item Concerning experimental aspects, we show that the 
FS's in the cubic phase will be useful in analyzing high-resolution 
Compton scattering and positron-annihilation measurements on the one
hand, and in verifying the present theoretical predictions 
on the other, suggesting the value of further experimental 
work along these lines. We comment also on the band renormalization 
in Ba$_{1-x}$K$_{x}$BiO$_{3}$ implied in the light 
of some recent photoemission experiments.
\end{enumerate} 

\acknowledgements
This work is supported by the US Department of Energy under contract
W-31-109-ENG-38 and the Academy of Finland, and benefited from the 
allocation of time at the NERSC, the Northeastern University Advanced 
Scientific Computation Center (NU-ASCC), the Center for Scientific 
Computing, Helsinki, and the Institute of Advanced Computing, Tampere, 
and a travel grant from NATO.

\begin{figure}
  \caption{
Fermi surface of cubic Ba$_{1-x}$K$_{x}$BiO$_{3}$
for $x=0.67$. (a) gives a 3D rendition, 
while (b) and (c) are two (001)-sections at $k_z=0$ and $k_z=\pi /2a$. 
(d) is a (110)-section through the zone center. The boundaries of the 
simple cubic zone (solid) as well as those of the bcc zone (dashed)
are shown. Points in (b)-(d) are the computed KKR-CPA values while the 
associated solid curves are the fits to the FS based on Eqs. 1 and 2 
discussed in the text. The disordered induced smearing of the FS is 
very small and is not shown.}
  \label{FS_cub}
\end{figure}
\begin{figure}
  \caption{
Same as the caption to Fig. 1 except $x=0.4$.
The arrow shows the spanning vector along (111) related to the 
Hume-Rothery instability to the orthorhombic phase discussed in the text.}

  \label{FS_htc}
\end{figure}
\begin{figure}
  \caption{
Same as the caption to Fig. 1 except $x=0.13$. The FS is seen to 
make contact with the Brillouin zone boundary around the X-points.}
  \label{FS_vhs}
\end{figure}  
\begin{figure}
  \caption{
Same as the caption to Fig. 1 except $x=0$, referring to BaBiO$_{3}$. }
  \label{FS_pure}  
\end{figure}
\begin{figure}
  \caption{
Band structures of BaBiO$_{3}$, KBiO$_{3}$ and BaPbO$_{3}$ along high-symmetry
directions are compared in the vicinity of the Fermi energy $E_F$ (dot-dashed
horizontal lines). }
  \label{ban}
\end{figure}
\begin{figure}
  \caption{
Selfconsistent KKR-CPA densities of states in the vicinity of the 
Fermi energy in (a) Ba$_{1-x}$K$_{x}$BiO$_{3}$ and 
(b) BaPb$_{x}$Bi$_{1-x}$O$_{3}$ over the composition 
ranges x indicated. Vertical dashed lines give Fermi energies.}
  \label{dos}
\end{figure}
\begin{figure}
  \caption{
Chemical potential $\mu(x)$ as a function of K concentration $x$ 
obtained from photoemission experiments by Ref. [34]
is seen to be in reasonable accord with KKR-CPA 
predictions renormalized by $0.49$ 
(dashed). Solid curve is the theoretical result based on Eq. 2.  }
  \label{mu}
\end{figure}

\begin{thebibliography}{99}
\bibitem{footnote1} Setting aside issues of resolution 
and statistics, and other practical matters tied to the fact that 
Compton measures a 2D integral and the positron annihilation a 1D or 
a 2D integral of $\rho({\mathbf p})$ and involves complications 
introduced by the positron wavefunction.
\end{thebibliography}
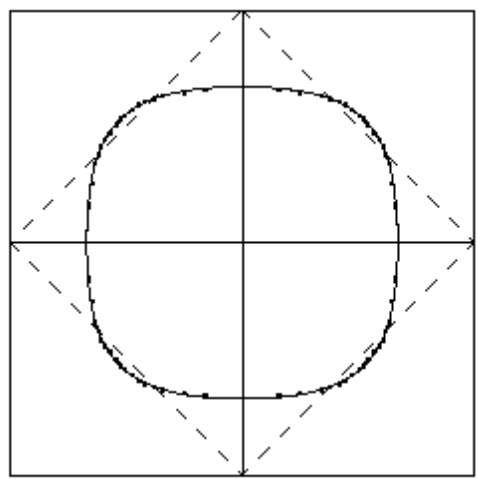
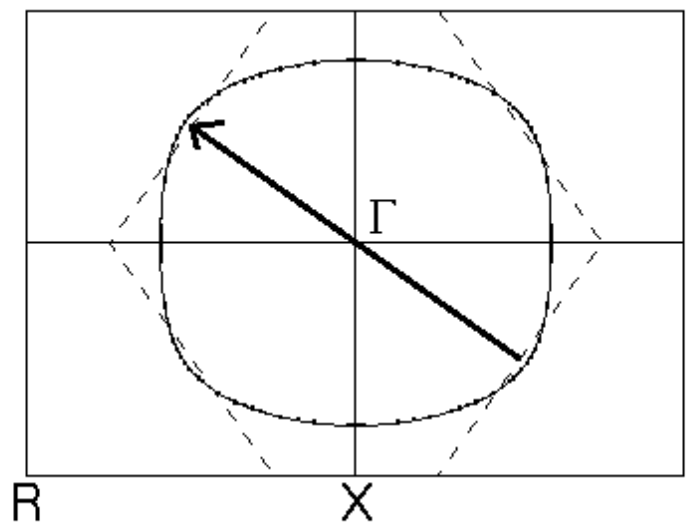
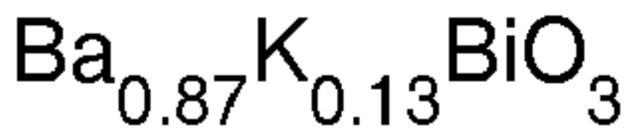
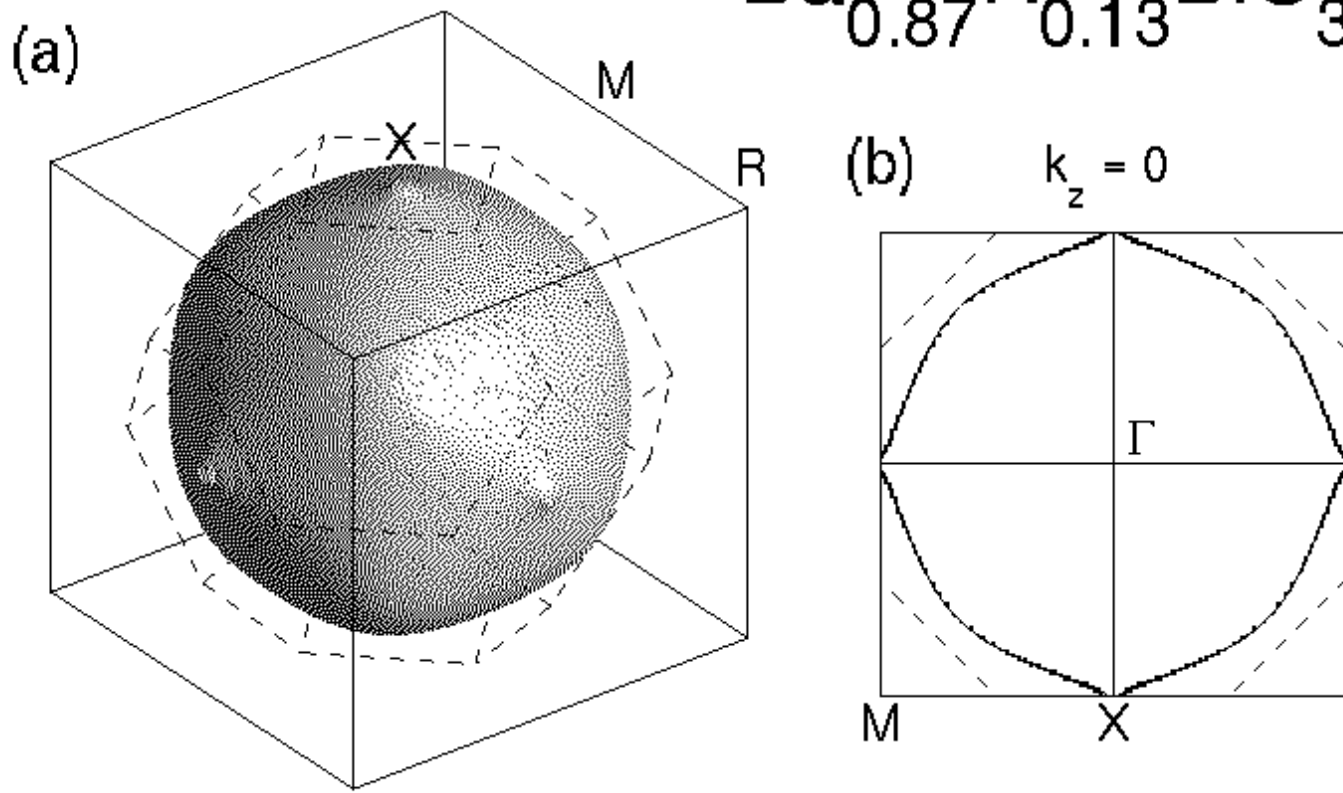
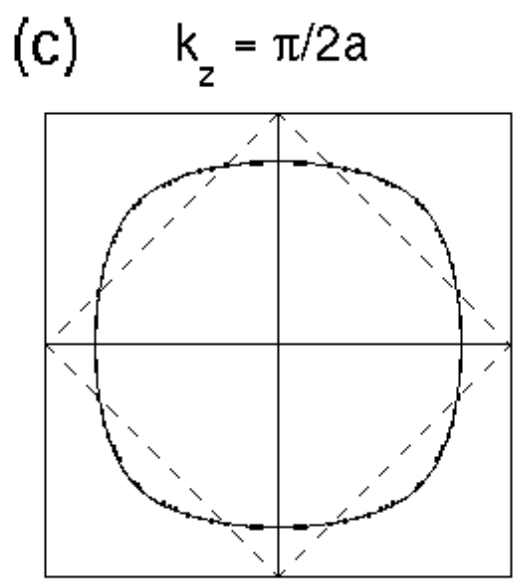
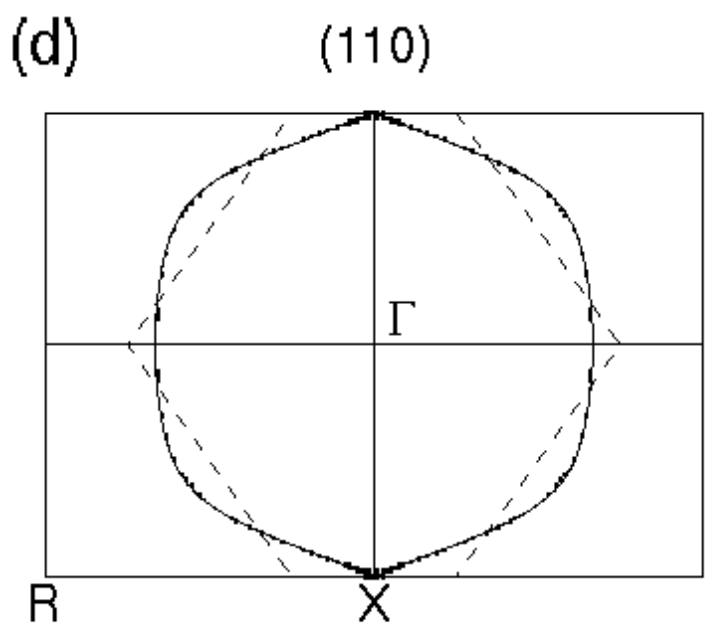
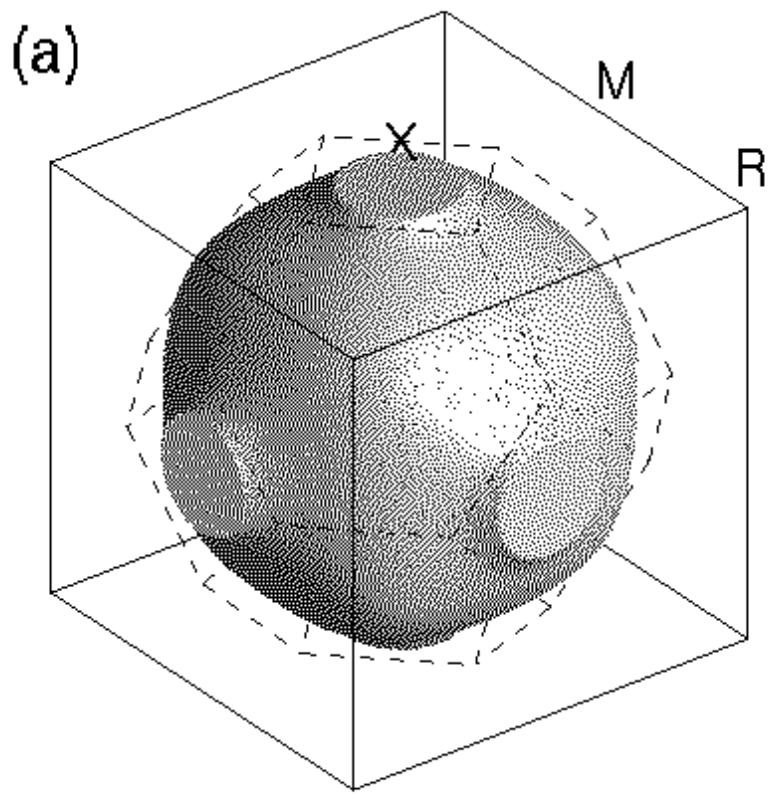
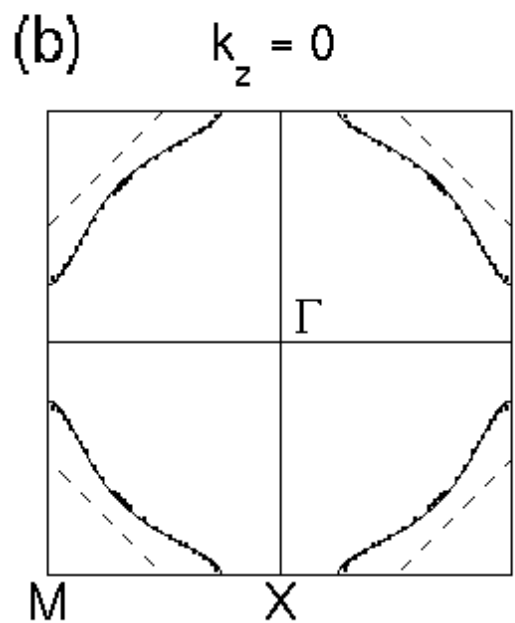
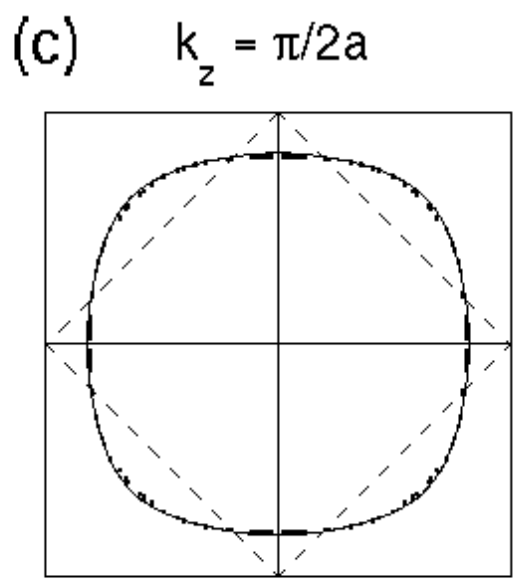
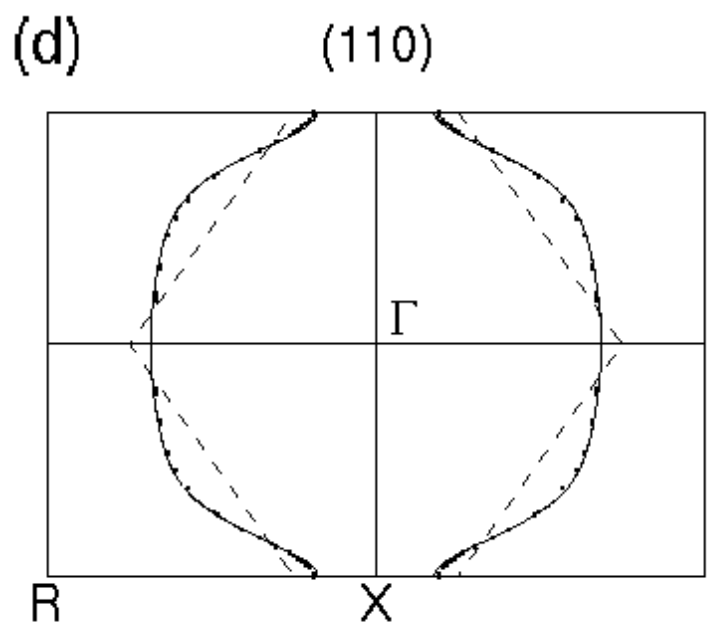
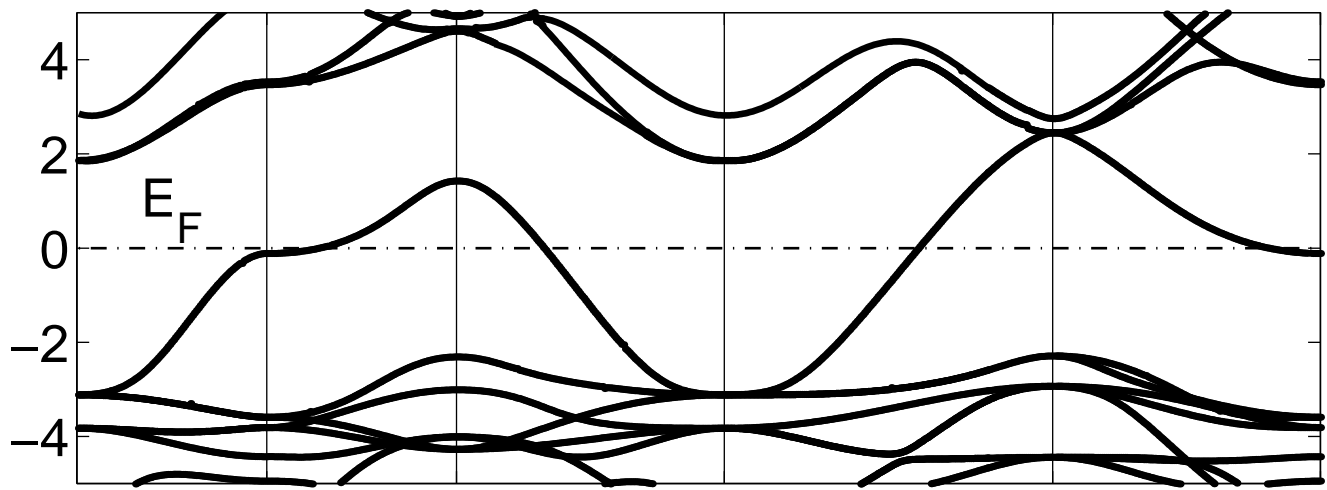
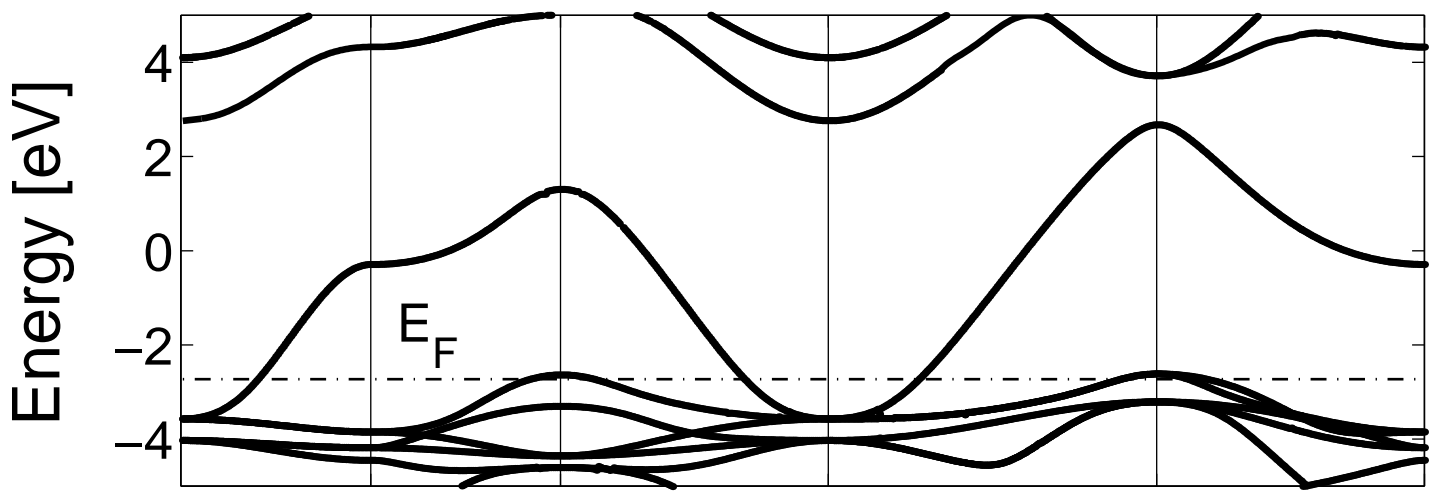
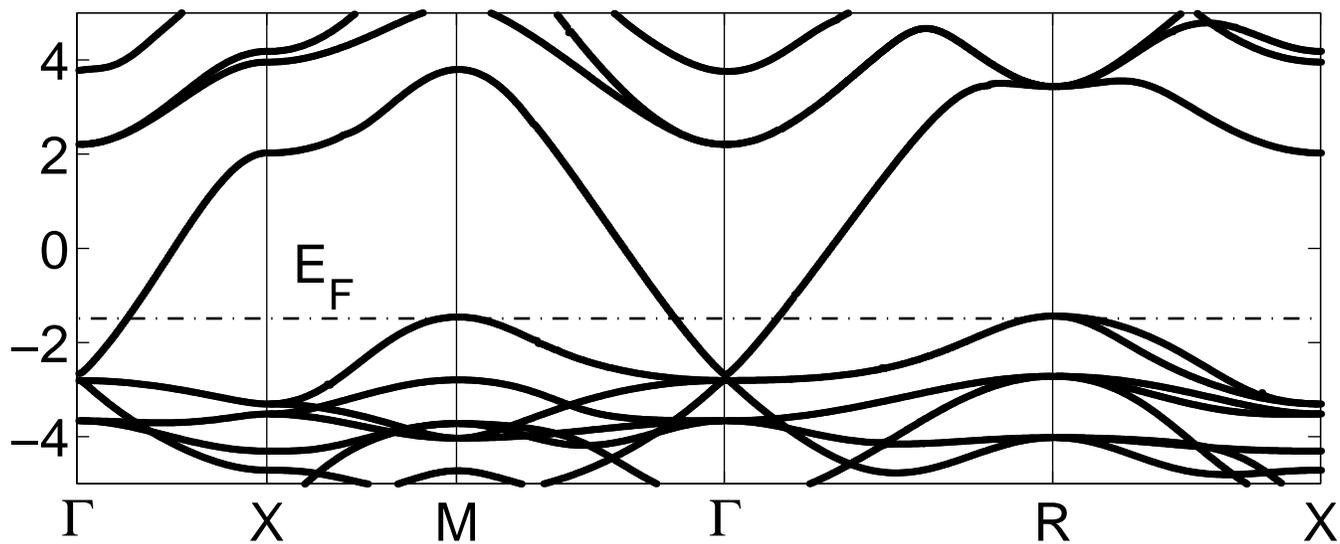
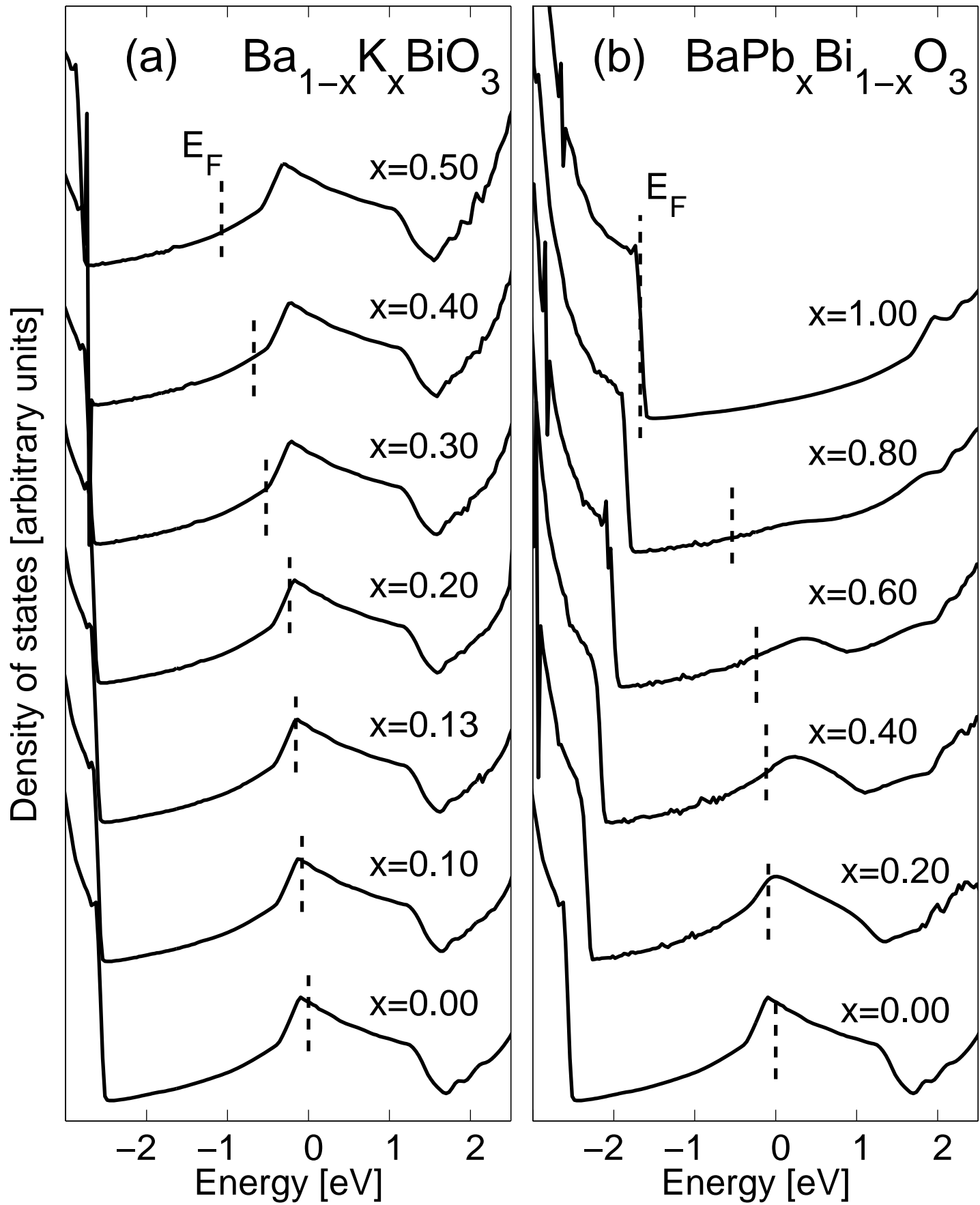
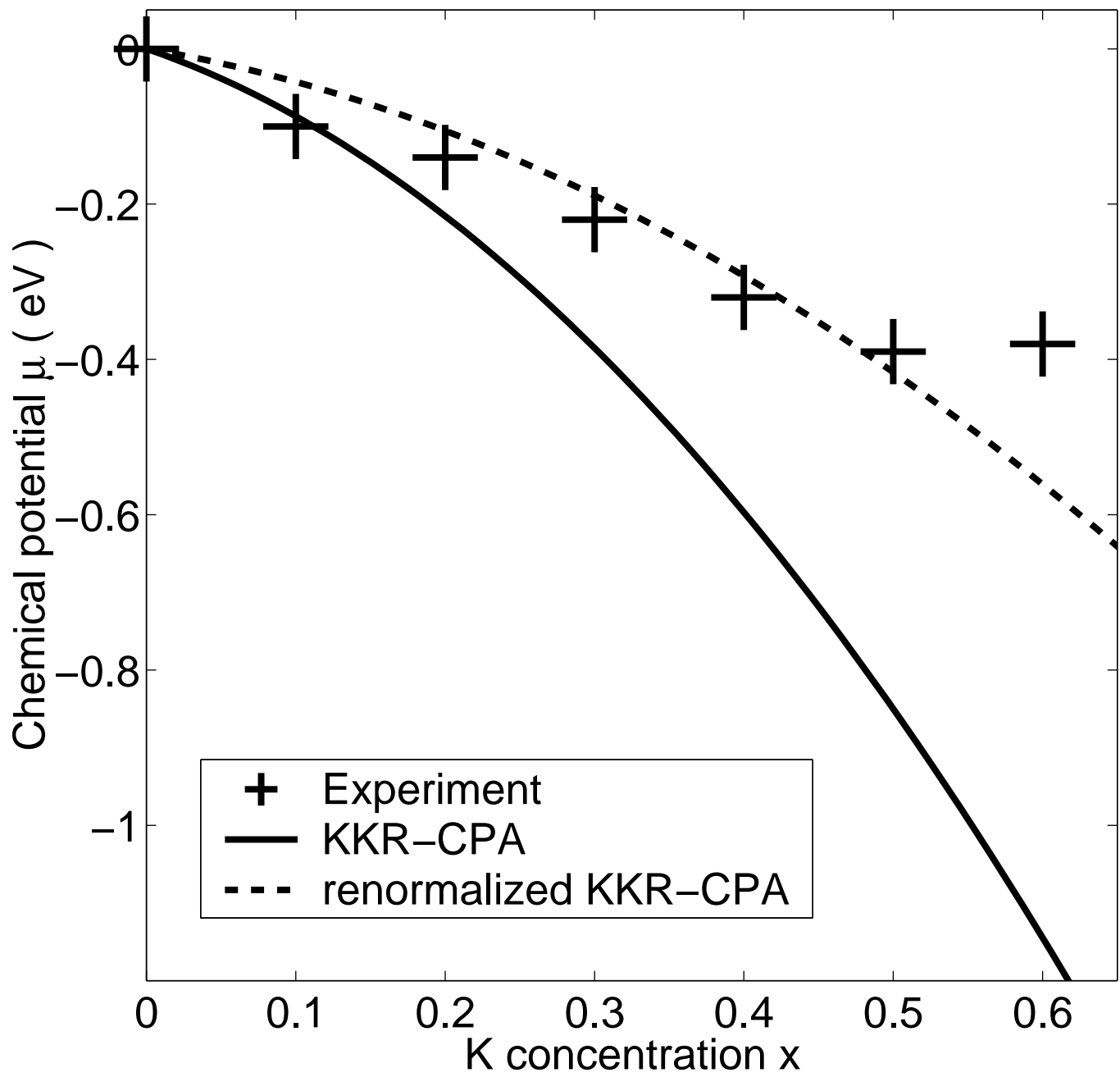
\end{document}